\documentclass[runningheads]{llncs}
\usepackage[utf8]{inputenc}
\usepackage{microtype}
\usepackage{amsmath,amssymb,amsfonts}
\usepackage{algorithm}
\usepackage{algpseudocode}
\usepackage[final]{pdfpages} 
\usepackage{booktabs}
\usepackage{multicol}
\usepackage{macro}

\DeclareMathOperator{\lcm}{lcm}
\newcommand{\unroll}[1]{\mathcal{U}(#1)}
\newcommand{\tree}[1]{\mathbf{\lowercase{#1}}}

\newcommand{\bigo}[1]{\mathcal{O}(#1)}

\newcommand{\ie}{\emph{i.e.}\@\xspace}

\newcommand{\setSize}[2]{(#1)(#2)}
\newcommand{\setDive}[2]{(#1)\{#2\}}

\usepackage[color=red!60,textsize=small]{todonotes}

\newcommand{\setLength}[1]{L(#1)}
\newcommand{\minLength}[1]{\ell(#1)}

\newcommand{\setDiv}[1]{\mathrm{Div}(#1)}
\DeclareMathOperator{\alcm}{alcm}
\newcommand{\aLcm}[2]{\alcm_{#2} #1}
\newcommand{\cycle}[1]{C_{#1}}
\newcommand{\qi}[1]{p_i^{#1}}
\newcommand{\qiai}{\qi{a_i}}
\newcommand{\qibi}{\qi{b_i}}

\newcommand{\prodC}[1]{\prod_{\substack{i=1\\#1}}^{n}}

\spnewtheorem{Theorem}{Theorem}{\bfseries}{\itshape}
\spnewtheorem{Lemma}[Theorem]{Lemma}{\bfseries}{\itshape}
\spnewtheorem{Corollary}[Theorem]{Corollary}{\bfseries}{\itshape}
\spnewtheorem{Proposition}[Theorem]{Proposition}{\bfseries}{\itshape}
\spnewtheorem{algo}[Theorem]{Algorithm}{\bfseries}{\itshape}
\spnewtheorem{Example}[Theorem]{Example}{\bfseries}{\itshape}

\title{Solving ``pseudo-injective'' polynomial equations over finite dynamical systems}
\titlerunning{Solving ``pseudo-injective'' polynomial equations over FDDS}
\author{
	Antonio E. Porreca
	\and
	Marius Rolland
}

\institute{
	Aix-Marseille Université, CNRS, LIS, Marseille, France\\
        \email{marius.rolland@lis-lab.fr}
}
\authorrunning{A.E. Porreca \and M. Rolland}

\date{\today}

\begin{document}
	\maketitle
	\begin{abstract}
        We consider the semiring of abstract finite dynamical systems up to isomorphism, with the operations of alternative and synchronous execution. We continue searching for efficient algorithms for solving polynomial equations of the form~$P(X)=B$, with a constant side~$B$, with the goal of decomposing complex behaviors into simpler systems. Taking inspiration from the characterization of injective polynomials~$P$ over dynamical systems, which is based on a condition on the lengths of limit cycles of their coefficients, we introduce a more general notion of pseudo-injectivity by relaxing this constraint. We prove that the associated equations can be solved efficiently, even in certain cases where the input is encoded in an exponentially more compact way.
	\end{abstract}
	
	\section{Introduction}
\label{sec:introduction}

Abstract finite discrete-time dynamical systems (FDDS) are pairs $(X,f)$ where $X$ is a finite set of states and $f \colon X \to X$ is a transition function (when~$f$ is implied, we denote $(X,f)$ simply as $X$). These objects abstractly represent the dynamic of concrete deterministic models such as finite cellular automata~\cite{parity_ca}, automata networks~\cite{automata_book} or reaction systems~\cite{reaction_systems}.

We often identify these systems with their transition digraphs, which have uniform out-degree one (Fig.~\ref{fig:product}).
These graphs are also known as functional digraphs. 
Their general shape is a collection of cycles, where each node of a cycle is the root of a finite in-tree (a directed tree with the edges pointing toward the root).
The nodes of the cycles are periodic states, while the others are transient states.

The set $(\mathbb{D}, +, \times)$ of FDDS up to isomorphism, with the alternative execution of two systems as \emph{addition} and the synchronous parallel 
execution of two FDDS as \emph{multiplication}, is a commutative semiring~\cite{article_fondateur}. 
As all semirings, we can define a semiring of polynomials over~$\mathbb{D}$ and the associated polynomial equations.

Although it has already been proven that general polynomial equations over~$\mathbb{D}$ (with variables on both sides of the equation) are undecidable, it is easily proved that, if one side of the equation is constant, then the problem becomes decidable (there is just a finite number of possible solutions)~\cite{article_fondateur}.
This variant of the problem is actually in $\NP$: since sums and products can be computed in polynomial time, we can just guess the values of the variables (their size is bounded by the constant side of the equation), evaluate the polynomial, and compare it with the constant side (isomorphism of functional digraphs can be checked in linear time~\cite{testIsoLinear}).

However, more restricted equations are not yet classified in terms of complexity.
For example, we do not know if monomial equations of the form $AX = B$ are $\NP$-hard, in $\P$, or possible candidates for an intermediate difficulty. 
However, it has been proved that we can find in polynomial time with respect to the size of the inputs ($A$, $B$ and~$\log k$) the (unique) connected solution of~$AX^k = B$ for all positive integer~$k$~\cite{kroot}.
It has also been shown that, when the polynomial $P = \sum_{i=0}^{m} A_i X^i$ is \emph{injective}, the equation $P(X) = B$ can be solved in polynomial time. The injective polynomials are exactly those where the coefficient~$A_i$ of a non-constant monomial (\ie, with~$i>1$) is \emph{cancelable} (Fig.~\ref{fig:product}), that is, it has a fixed point~\cite{poly_inj}.
We are interested in finding a generalization of this result to a larger class of polynomials.

We can reformulate the condition of injectivity by saying that there exists an~$i \ge 1$ such that $A_i$ is cancelable (\ie, $A_i B = A_i C$ if and only $B = C$ for all $B,C$)~\cite{poly_inj}.
If an FDDS $A$ is cancelable, then it satisfies the following property: denoting by $\setLength{A}$ the set of all cycle lengths in $A$, it is always the case that $\min(\setLength{A})$ divide all elements of $\setLength{A}$.
We call \emph{pseudo-cancelable} a element which verifies this property, without being necessarily cancelable (Fig.~\ref{fig:product}).

In this paper we first prove that equations of the form~$AX=B$ can be solved in polynomial time if~$A$ is pseudo-cancelable, then generalize this result to an efficient algorithm for equations of the form~$P(X)=B$ where~$P = \sum_{i=0}^{m} A_i X^i$ is a polynomial such that~$\sum_{i=1}^{m} A_i$ (\ie, the sum of coefficient of non constant terms) is pseudo-cancelable. We refer to this kind of polynomials as \emph{pseudo-injective}.



	\section{Definitions}

The set of FDDS \emph{up to isomorphism}, with mutually
exclusive alternative execution as \emph{sum} and synchronous parallel execution as \emph{product}, forms a semiring~\cite{article_fondateur}, denoted by $\mathbb{D}$. 
The two operations above define a notion of algebraic decomposition of FDDS.

\begin{figure}[t]
\centering
\includegraphics[page=2,scale=1.2]{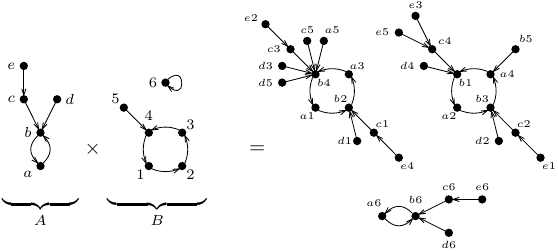}
\caption{Product of two FDDS~$A$ and~$B$, where the states are given a temporary label in order to show how the result is computed. Remark that~$B$ is cancelable (state $6$~is a fixed point) and~$AB$ is pseudo-cancelable ($\setLength{AB} = \{2,4\}$ and~$\minLength{AB}=2$). $A$~is also trivially pseudo-cancelable, since it is connected ($\setLength{A} = \{2\}$ and~$\minLength{A}=2$).}
\label{fig:product}
\end{figure}

The semiring of function digraphs, with disjoint union for sum and direct product for multiplication, is isomorphic to $\mathbb{D}$.
We recall that the direct product of two graphs $A,B$ is the graph $C$
such that $V(C) = V(A) \times V(B)$ and
$E(C) = \{((u,u'), (v,v')) \mid (u,v) \in E(A), (u',v') \in E(B)\}$~\cite{livreGraphe}. 
Then, an FDDS can be seen as the sum of connected components, where each component consists of a cycle, representing the periodic behavior, and in-trees (trees directed from leaf to root) rooted in the cycle, representing the transient behavior. An example of the product of two FDDS is depicted in Fig.~\ref{fig:product}.

The structure of the product is algebraically rich.
For example, this semiring is not factorial, \ie, there exist four irreducible FDDS $A,B,C,D$ such that $A \neq C$ and $A \neq D$ but $AB = CD$.
Moreover, while the periodic behavior of a product is easy to analyze, its transient behavior demands much more work.
Indeed, the product of two connected FDDS $A,B$, having cycles of length~$p$ and~$q$ respectively,
generates $\gcd(p,q)$ connected components, each with a cycle of size $\lcm(p,q)$~\cite{livreGraphe}.
In order to analyze the (generally more complex) transient behaviors, we use the notion of unroll introduced in~\cite{article_arbre}.

\begin{definition}[Unroll]
	Let $A = (X,f)$ be an FDDS.
	For each state $u\in X$ and $k \in \mathbb{N}$, we denote by $f^{-k}(u) = \{ v \in X \mid f^k(v) = u \}$ the set of $k$-th preimages of $u$. 
	For each $u$ in a cycle of $A$, we call the \emph{unroll tree of $A$ in $u$} the infinite tree $\tree{t}_u = (V,E)$ having vertices $V = \{(s,k) \mid s \in f^{-k}(u), k \in \mathbb{N}\}$ and edges
	$E = \big\{ \big((v,k),(f(v),k-1) \big) \big\} \subseteq V^2$.
	We call \emph{unroll of $A$}, denoted $\unroll{A}$, the set of its unroll trees.
\end{definition}

An unroll tree contains exactly one infinite branch, onto which are periodically rooted the trees representing the transient behaviors of the corresponding connected component.
Remark that the set of unrolls \emph{up to isomorphism} (which will be used in the rest of this paper, unless otherwise specified) with the disjoint union as sum and the levelwise product defined in~\cite{article_arbre} (and already exploited in~\cite{kroot,poly_inj}) is also a semiring, denoted by $\mathbb{U}$.
In addition, the operation of unrolling an FDDS is a semiring homomorphism $\mathbb{D} \to \mathbb{U}$.  

An interesting aspect of $\mathbb{U}$ is the existence of a total order compatible with the product~\cite{article_arbre}, that is, for three unroll trees $\tree{t}_1, \tree{t}_2, \tree{t}$ we have $\tree{t}_1 \le \tree{t}_2$ if and only if $\tree{t}_1 \tree{t} \le \tree{t}_2 \tree{t}$. 
This features has been exploited in order to describe polynomial algorithms for solving certain classes equations of the form $A X = B$ or $P(X) = B$ with specific condition on $A,X, P$ or $B$~\cite{article_arbre,kroot,poly_inj}.

Remark that an order compatible with the product does not exist on FDDS. 
Indeed, if such an order existed then all FDDS would be cancelable;
however, the cancelable elements in $\mathbb{D}$, are exactly the FDDS that admit 
a fixed point~\cite[Theorem 30]{article_arbre}. 

In this article we focus our attention on the lengths of the cycles of an FDDS. Denote by $\setLength{A}$ the set of cycle lengths of a connected component of $A$, and denote $\minLength{A} = \min(\setLength{A})$.
Since least common multiples are important when dealing with the product of connected components, it is useful to introduce the following notion of ``anti-lcm'':

\begin{definition}[anti-lcm] Let $a, b$ be positive integers with factorizations into primes~$\prod_{i=1}^{n} p_i^{a_i}$ and~$\prod_{i=1}^{n} p_i^{b_i}$ respectively and such that~$a$ divides~$b$. Then, the \emph{anti-lcm of~$b$ with respect to~$a$} is \[\aLcm{b}{a} = \prodC{b_i > a_i} p_i^{b_i}.\]
\end{definition}

That is, $\aLcm{b}{a}$ is the product of the primes in the factorizations of~$a$ and~$b$ with their exponents in~$b$, but only if they are strictly larger than their exponents in~$a$. For instance, if~$a = 12 = 2^2 \times 3^1$ and~$b = 252 = 2^2 \times 3^2 \times 7^1$ then~$\aLcm{b}{a} = 3^2 \times 7^1 = 63$. An important aspect of this notion (which justifies the name ``anti-lgm'') is that the anti-lcm of $\lcm(a,b)$ with respect to $a$ is the smallest integer $c$ such that $\lcm(a, c) = \lcm(a,b)$.

From the characterization of cancelable elements of $\mathbb{D}$, we remark that if $A$ is cancelable then every $a \in \setLength{A}$ is a multiple of $\minLength{a}$. 
By dropping the cancelability requirement, we say that an FDDS $A$ is \emph{pseudo-cancelable} if every $a \in \setLength{A}$ is multiple of $\minLength{a}$ (Fig.~\ref{fig:product}).
Moreover, we call a polynomial $P = \sum_{i=0}^{m} A_i X_i$ over~$\mathbb{D}$ \emph{pseudo-injective} if the sum of the coefficients of its non-constant terms~$\sum_{i=1}^{m} A_i$ is pseudo-cancelable; notice that this is indeed a generalization of the actual injectivity condition~\cite{poly_inj}.


	\section{The complexity of solving polynomial equations}\label{section:division}

The main goal of this section is to prove the following theorem:

\begin{Theorem}\label{th:sol_poly_is_poly}
	Let $B$ be an FDDS and $P = \sum_{i=0}^{m} A_i X_i$ be a pseudo-injective polynomial. 
	Then, we can solve $P(X) = B$ in polynomial time with respect to the size (number of states) of~$B$ plus the sizes of the coefficients of~$P$.
\end{Theorem} 

We begin by proving that this is true for equations $A X = B$ where $A,X$ and~$B$ are sums of cycles (\ie, their underlying transition functions are permutations).
To this end, we prove that every cycle in $X$ must satisfy a certain condition on its size.
For brevity, in the following we denote by $\cycle{i}$ the cycle with length $i$; remark that all sums of cycles~$A$ can be written as~$A = \sum_{i \in \setLength{A}} a_i C_i$, where~$a_i$ is the number of occurrences of~$C_i$ in~$A$. We also denote by~$\setDiv{p}$ the set of divisors of~$p$.

\begin{Lemma}\label{lemma:first_cond_cycle}
	Let~$A$, $B$ and~$X$ be sums of cycles with~$A$ pseudo-cancelable. \linebreak
	If~$A X = B$ then there exists a divisor~$k$ of $\minLength{A}$ satisfying $k \times \aLcm{A}{B} \in \setLength{X}$ and $\gcd(k, \aLcm{A}{B}) = 1$.
\end{Lemma}

\begin{proof}
	Since $A X = B$, by definition of product there exist $a \in \setLength{A}$ and $x \in \setLength{X}$ such that $\minLength{B} = \lcm(a,x)$. 
	Furthermore, since~$B$ contains all products of cycles from~$A$ with cycles from~$X$, in particular we have $\lcm(\minLength{A}, x) \in \setLength{B}$.
	It follows that $\lcm(\minLength{A},x) \geq \minLength{B}$.
	Since $A$ is pseudo-cancelable, necessarily $\minLength{A}$ divides~$a$ and thus $\lcm(\minLength{A}, x)$ divides $\lcm(a, x)$. 
	Then $\lcm(\minLength{A},x)~\le~\lcm(a,x)~=~\minLength{B}$ and therefore $\lcm(\minLength{A},x) = \minLength{B}$. This implies that the primes in the factorization of~$\minLength{B}$ whose exponent is larger than in~$\minLength{A}$ must belong to the factorization of~$x$. Hence, $\aLcm{A}{B}$ divides~$x$ and therefore $x = k \times \aLcm{A}{B}$ for some integer~$k$. Since~$x$ divides~$\minLength{B}$, no prime in~$k$ appears in~$\aLcm{A}{B}$ (otherwise their exponent in~$x$ would be larger than in~$b$) and thus $\gcd(k, \aLcm{A}{B}) = 1$ and~$k$ divides~$\minLength{B} / \aLcm{A}{B}$. Finally, \[\minLength{B} / \aLcm{A}{B} = \prodC{a_i=b_i} p_i^{a_i}\] and thus this number divides~$\minLength{A}$, implying that~$k$ is also a divisor of~$\minLength{A}$.
\end{proof}

\begin{Corollary}\label{cor:fix_min}
	If $AX = B$, where~$A$ and~$B$ are two sum of cycles and $A$ is pseudo-cancelable, then for each cycle $\cycle{x} \in X$ which verifies $\cycle{\minLength{B}} \in A \cycle{x}$ there exists a divisor~$k$ of $\minLength{A}$ satisfying $k \times \aLcm{A}{B} = x$ and $\gcd(k, \aLcm{A}{B}) = 1$.
\end{Corollary}

Although Lemma~\ref{lemma:first_cond_cycle} and Corollary~\ref{cor:fix_min} partially describe a cycle of $X$, there still remains an unknown, namely the value of $k$. 
Nevertheless, as we show in the following lemma, we can fix $k=1$.

\begin{Lemma}\label{lemma:second_cond_cycle}
	Let~$A = \sum_{i \in \setLength{A}} a_i \cycle{i}$ and~$B, X$ be sums of cycles with~$A$ pseudo-cancelable and let $k \in \setDiv{\minLength{A}}$ such that $\gcd(k, \aLcm{A}{B}) = 1$.
	Then, we have~$A (X + \cycle{k \times \aLcm{A}{B}}) = B$ if and only if $A (X + d \cycle{(k/d) \times \aLcm{A}{B}}) = B$ for all~$d \in \setDiv{k}$.
\end{Lemma} 

\begin{proof}
	Let $x = k \times \aLcm{A}{B}$, $d \in \setDiv{k}$ and $a \in \setLength{A}$.
	Since~$A$ is pseudo-cancelable, we have $\minLength{A} \in \setDiv{a}$. 
	Furthermore, from the definition of $k$, we deduce that $k \in \setDiv{a}$ and also $d \in \setDiv{a}$.
	Consequently, 
	\[\gcd(x,a) = \gcd(\aLcm{A}{B}, a) \gcd(k,a) = k \times \gcd(\aLcm{A}{B}, a).\]
	In addition, we have~$k = \gcd(k,a) = \gcd(d \times (k/d), a) = d \times \gcd(k/d,a)$ and $\lcm(x,a) = \lcm(\aLcm{A}{B}, a)$.
	By consequence,
	\begin{align*}
		\cycle{a} \cycle{x} &= \gcd(x,a) \cycle{\lcm(x,a)} \\
		&= k \times \gcd(\aLcm{A}{B}, a) \times \cycle{\lcm(\aLcm{A}{B}, a)} \\
		&= d \times \gcd(k/d,a) \times \gcd(\aLcm{A}{B}, a) \times \cycle{\lcm(\aLcm{A}{B}, a)} \\
		&= d \times \gcd((k/d) \times \aLcm{A}{B}, a) \times \cycle{\lcm((k/d) \times \aLcm{A}{B}, a)}\\
		&= \cycle{a} \times d \cycle{\lcm((k/d) \times \aLcm{A}{B}, a)}
	\end{align*}
	This implies
	\begin{align*}
		A \cycle{x} &= \sum_{i \in \setLength{A}} a_i \cycle{i} \cycle{x}\\ 
		&= \sum_{i \in \setLength{A}} a_i d \cycle{i} \cycle{\lcm((k/d) \times \aLcm{A}{B}, a)}\\
		&= A d \cycle{\lcm((k/d) \times \aLcm{A}{B}, a)}
	\end{align*}
	and the lemma follows.
\end{proof}

We can now formulate an algorithm for the resolution of $AX = B$ working under the hypothesis that $A$ and $B$ are sums of cycles and $A$ is pseudo-cancelable.

\begin{algo}\label{algo:div_cycle}
	Given two sums of cycles~$A$ and $B$ with $A$ pseudo-cancelable, we can compute $X$ such that $AX=B$, if any exists, in the following way:
	\begin{enumerate}
		\item \label{algo_div_cycle:step1} Check if $|A| \le |B|$. 
		\item\label{algo_div_cycle:step2}  Let $C$ be a cycle of length $\aLcm{A}{B}$.
		\item\label{algo_div_cycle:step3} Check if $AC$ is a submultiset of $B$.
		\item\label{algo_div_cycle:step4} Add $C$ to $X$.
		\item\label{algo_div_cycle:step5} Set $B \gets B-AC$.
		\item\label{algo_div_cycle:step6} If $B \neq \emptyset$ then go to \ref{algo_div_cycle:step1}; otherwise return $X$.
	\end{enumerate}
\end{algo}

This algorithm runs in polynomial time with respect to the number of states of $A$ and $B$. 
In fact, the number of iterations is bounded by $n = |B|$. 
Moreover, all the instructions in the loop only require polynomial time.
Indeed, line~\ref{algo_div_cycle:step1} is~$\bigo{n}$ time.
Then, we can compute $\aLcm{A}{B}$ simply by counting up to $\minLength{B}$ and construct $C$ in $\bigo{n}$ time, since $n \ge \aLcm{A}{B}$.
Thus, line~\ref{algo_div_cycle:step2} is executed in $\bigo{n}$ time.
Then, we can compute $AC$ in $\bigo{n^2}$ time, and its size is bounded by $n^2$. 
It follows that checking if $AC$ is a submultiset of $B$ and removing $AC$ from $B$ requires $\bigo{n^3}$ time.
We deduce that lines~\ref{algo_div_cycle:step3} and~\ref{algo_div_cycle:step5} are executed in~$\bigo{n^3}$ time.
All the other instructions only require constant time.

Furthermore, the solution $X$ computed by Algorithm~\ref{algo:div_cycle} is correct. 
Indeed, if~$m$ is the number of iterations of the algorithm, $C_i$ the cycle computed in the $i$-th iteration and $B_i$ the value of $B$ at the beginning of the $i$-th iteration, we have \mbox{$AC_m = B_m$}. 
Inductively, we have $B_i = A C_i + B_{i+1}$. 
Since~\mbox{$B_{i+1} = A \sum_{j = i+1}^{m} C_j$}, it follows that $B_i = A \sum_{j=i}^{m} C_j$.

All that remains to prove in order to show the correctness of the algorithm is that, if a solution exists, then the algorithm finds it.

\begin{Lemma}\label{lemma:correcteness_algo_div_cycle}
	Let $A$ and $B$ be sums of cycles with $A$ pseudo-cancelable. If the equation~$AX=B$ has solutions, then Algorithm~\ref{algo:div_cycle} finds one maximizing the number of connected components (equivalently, minimizing the cycle lengths).
\end{Lemma}

\begin{proof}
	Suppose that $AX = B$. 
	Thus, by Lemma~\ref{lemma:first_cond_cycle}, there exist $k_1$ and~$X_1$ such that $X = X_1 + C$ with $C = \cycle{k_1 \times\aLcm{A}{B}}$. 
	Accordingly, we have $A X_1 = B_1$ with $B_1 = B - A C$.
	Consequently, by Lemma~\ref{lemma:second_cond_cycle}, the sum of cycles $Y = X_1 + k_1 \cycle{\aLcm{A}{B}}$ satisfies $A Y = B$.
	This implies $B_1 = B - A k_1 \cycle{\aLcm{A}{B}}$.
	
	The number of cycles with length $\minLength{B}$ generated by $A C$ is 
	\[\sum_{\substack{a \in \setLength{A}\\a \mid \minLength{B}}} \gcd(\aLcm{A}{B}, a) k_1\] 
	whereas the number generated by $A \cycle{\aLcm{A}{B}}$ is
	\[\sum_{\substack{a \in \setLength{A}\\a \mid \minLength{B}}} \gcd(\aLcm{A}{B}, a).\] 
	We deduce that the value of $\minLength{B}$ in line~\ref{algo_div_cycle:step2} of the algorithm is the same between the first iteration and the $k_1$th iteration of the loop. 
	This implies that the result contains $k_1 \cycle{\aLcm{A}{B}}$ at the end the $k_1$th iteration of the loop.
	
	The lemma follows by induction. Remark that the solution~$Y$ constructed at each step replaces one cycle by the maximum number of cycles of minimum compatible length while still satisfying the equation; this implies that the solution found by the algorithm maximizes the number of connected components
\end{proof}


This proves the correctness of~Algorithm~\ref{algo:div_cycle}:

\begin{Proposition}\label{prop:div_cycle_poly}
	Given two sums of cycles~$A$ and $B$ with $A$ pseudo-cancelable, we can solve in polynomial time the equation $AX=B$.
\end{Proposition}

We can now extend our result to the case of equations $P(X) = B$ where~$B$ and all coefficients of~$P$ are sums of cycles.
First, we generalize Lemma~\ref{lemma:first_cond_cycle} as follows:

\begin{Lemma}\label{lemma:first_cond_cycle_poly}
	Let $B$ and $Y$ be two sums of cycles and $P = \sum_{i=0}^{m} A_i X^i$ be a pseudo-injective polynomial where each coefficient is a sum of cycles. 
	Let $A_c$ be a coefficient of $P$ such that $\minLength{A_c} = \minLength{\sum_{i=1}^{m} A_i}$. 
	Then $P(Y) = B$ implies the existence of~$k \in \setDiv{\minLength{A_c}}$ such that $k \times \aLcm{A_c}{B-A_0} \in \setLength{Y}$ and $\gcd(k, \aLcm{A_c}{B-A_0}) = 1$.
\end{Lemma}

\begin{proof}
	Suppose $P(Y) = B$.
	Thus, $P(Y) - A_0 = B'$ with $B' = B - A_0$.
	Therefore, there exist $i \in \{1,\ldots,m\}$, $a \in \setLength{A_i}$ and $y \in \setLength{Y}$ such that $\minLength{B'} = \lcm(a,y)$.
	Furthermore, for all $j \in \{1,\ldots,m\}$ we have $\lcm(\minLength{A_j}, y) \in \setLength{B'}$.
	It follow that $\lcm(\minLength{A_j},y) \geq \minLength{B'}$ for all $j$.
	By hypothesis on $P$, $\minLength{A_c}$ divides $a$. 
	Thus, $\lcm(\minLength{A_c},y) \le \lcm(a,y) = \minLength{B'}$.
	We deduce that $\lcm(\minLength{A_c},y) = \minLength{B'}$ and therefore $y = k \aLcm{A_c}{B'} $ with $k \in \setDiv{\minLength{A_c}}$ and $\gcd(k, \aLcm{A_c}{B'}) = 1$.
\end{proof}

Thus, we can generalize Lemma~\ref{lemma:second_cond_cycle} as follows.

\begin{Lemma}\label{lemma:second_cond_cycle_poly}
	Let $B, Y$ be two sums of cycles and $P = \sum_{i=0}^{m} A_i X^i$ be a pseudo-injective polynomial where each coefficient is a sum of cycles. 
	Let $A_c$ be a coefficient of $P$ such that $\minLength{A_c} = \minLength{\sum_{i=1}^{m} A_i}$. 
	Let $k \in \setDiv{\minLength{A_c}}$ such that $k \times \aLcm{A_c}{B-A_0} \in \setLength{Y}$ and $\gcd(k, \aLcm{A_c}{B-A_0}) = 1$. Then, we have
	$P(Y + \cycle{k \times \aLcm{A_c}{B}}) = B$ if and only if  $P(Y + d \cycle{(k/d) \times \aLcm{A_c}{B}}) = B$ for all~$d \in \setDiv{k}$.
\end{Lemma} 

\begin{proof}
	Let $y = k \times \aLcm{A_c}{B}$, $d \in \setDiv{k}$ and $a \in \setLength{\sum_{i=1}^{m} A_i}$.
	Then, we have~$\minLength{A_c} \in \setDiv{a}$. 
	Furthermore, from the definition of $k$, we deduce that $k$ and $d$ divide~$a$.
	Consequently, from the proof of Lemma~\ref{lemma:second_cond_cycle}, it follows that
	\[\cycle{a} \cycle{y}^j = \cycle{a} \cycle{y}^{j-1} (d \cycle{(k/d) \times \aLcm{A_c}{B}}) = \cycle{a} (d \cycle{(k/d) \times \aLcm{A_c}{B}})^j\]
	for all natural numbers $j$.
	Thus by summing, $A_j \cycle{y}^j = A_j (d \cycle{(k/d) \times \aLcm{A_c}{B}})^j$ for all $j$. 
	We conclude that $P(Y + \cycle{y}) = P(Y + d \cycle{(k/d) \times \aLcm{A_c}{B}})$.
\end{proof}

We deduce that an adaptation of Algorithm~\ref{algo:div_cycle} can solve in polynomial time the equation $P(Y) = B$ if $P$ is a pseudo-injective polynomial (Fig.~\ref{fig:run-algo} shows an example run).

\begin{algo}\label{algo:poly_cycle}
	Given a sum of cycles $B$ and a pseudo-injective polynomial $P = \sum_{i=0}^{m} A_i X^i$ where each coefficient is a sum of cycles, we can compute $Y$ such that $P(Y)=B$, if any exists, in the following way:
	\begin{enumerate}
		\item Check if~$A_0$ is a submultiset of~$B$.
		\item Set $B \gets B - A_0$, $P \gets P - A_0$ and~$Y \gets \mathbf{0}$.
		\item\label{algo_poly_cycle:step2}  Let $C$ be a cycle of length $\aLcm{(\sum_{i=1}^{m} A_i)}{B}$.
		\item\label{algo_poly_cycle:step3} Check if $|P(Y+C)| = \sum_{i=1}^m |A_i||Y+C|^i$ is bounded by~$|B|$.\footnote{Remark that $|A + B| = |A| + |B|$ and $|A \times B| = |A| \times |B|$ for all FDDS~$A, B$ by the definitions of sum and product (in other words, the size function $|\cdot| \colon \mathbb{D} \to \mathbb{N}$ is a semiring homomorphism).}
		\item\label{algo_poly_cycle:step4} Compute~$P(Y+C)$ and check if $P(Y + C) - P(Y)$ is a submultiset of $B$.
		\item\label{algo_poly_cycle:step5} Set $B \gets B - (P(Y + C) - P(Y))$ and $Y \gets Y + C$.
		\item\label{algo_poly_cycle:step7} If $B \neq \emptyset$ then go back to \ref{algo_poly_cycle:step2}, otherwise return $Y$.
	\end{enumerate}
\end{algo}


\begin{figure}[t]
	\centering
	\fontsize{8}{10}\selectfont
	\begin{tabular}{cc@{\hspace{0.5em}}c@{\hspace{0.5em}}@{\hspace{1em}}c@{\hspace{1em}}cc}
		\toprule
		$B$ & $Y$ & $C$ & $P(Y+C)$ & $P(Y)$ & $P(Y+C)-P(Y)$ \\
		\midrule
		$\begin{gathered}16C_2+4C_4+\\[-0.4em]18C_6+C_{12}\end{gathered}$ & $\mathbf{0}$ & $C_1$ & $C_2+C_4+C_6$ & $\mathbf{0}$ & $C_2+C_4+C_6$\\[1.5em]
		$\begin{gathered}15C_2+3C_4+\\[-0.4em]17C_6+C_{12}\end{gathered}$ & $C_1$ & $C_1$ & $4C_2+2C_4+2C_6$ & $C_2+C_4+C_6$ & $3C_2+C_4+C_6$\\[1.5em]
		$\begin{gathered}12C_2+2C_4+\\[-0.4em]16C_6+C_{12}\end{gathered}$ & $2C_1$ & $C_1$ & $9C_2+3C_4+3C_6$ & $4C_2+2C_4+2C_6$ & $5C_2+C_4+C_6$\\[1.5em]
		$\begin{gathered}7C_2+C_4+\\[-0.4em]15C_6+C_{12}\end{gathered}$ & $3C_1$ & $C_1$ & $16C_2+4C_4+4C_6$ & $9C_2+3C_4+3C_6$ & $7C_2+C_4+C_6$\\[1em]
		$14C_6+C_{12}$ & $4C_1$ & $C_3$ & $\begin{gathered}16C_2+4C_4+\\[-0.4em]18C_6+C_{12}\end{gathered}$ & $16C_2+4C_4+4C_6$ & $14C_6+C_{12}$\\[1em]
		$\mathbf{0}$ & $4C_1 + C_3$ &&&&\\
		\bottomrule
	\end{tabular}
	\caption{A run of Algorithm~\ref{algo:poly_cycle} on equation~$C_2X^2 + (C_4+C_6)X = 16C_2+4C_4+18C_6+C_{12}$, where each line of the table corresponds to an iteration of lines~\ref{algo_poly_cycle:step2}--\ref{algo_poly_cycle:step7}. This run gives the solution~$Y = 4C_1 + C_3$. Remark that this equation also has the solutions~$2C_2+C_3$ and~$2C_1+C_2+C_3$, but Algorithm~\ref{algo:poly_cycle} finds the solution which maximizes the number of connected components.}
	\label{fig:run-algo}
\end{figure}


Notice that the test in step~\ref{algo_poly_cycle:step3} is needed, since~$m$ can in principle be larger than~$|B|$ and then~$|Y+C| > 1$ would imply~$|P(Y+C)| \ge 2^m \ge 2^{|B|}$, which would demand exponential computation time. The test itself must be performed by using fast exponentiation and stopping as soon as an intermediate result exceeds~$|B|$ in order to guarantee polynomial time.
It follows that steps~\ref{algo_poly_cycle:step3},~\ref{algo_poly_cycle:step4} and~\ref{algo_poly_cycle:step5} require polynomial time with respect to the size of the input.
We conclude that this algorithm is polynomial-time.
In addition, as for Algorithm~\ref{algo:div_cycle}, if the algorithm returns~$Y$ then $P(Y) = B$. 
Finally, we can extend Lemma~\ref{lemma:correcteness_algo_div_cycle} by exploiting Lemmas~\ref{lemma:first_cond_cycle_poly} and~\ref{lemma:second_cond_cycle_poly} (instead of Lemmas~\ref{lemma:first_cond_cycle} and~\ref{lemma:second_cond_cycle}) and by replacing the products by~$A$ with evaluations of~$P$ for the other direction. 

\begin{Proposition}
	\label{prop:poly_cycle_polytime}
	Given a sum of cycles $B$ and a pseudo-injective polynomial  $P=\sum_{i=0}^{m} A_i X^i$ where each coefficient is a sum of cycles, we can find $Y$ such that $P(Y)=B$, if one exists, in polynomial time with respect to the number of states of the input.
\end{Proposition}

We can now consider the case where $A_0,\ldots,A_m$ and $B$ are not just sums of cycles. As in the previous case, our main idea is to iteratively construct the solution. 
At this end, we recall the notation introduced in~\cite{poly_inj}, namely, the function~$\setSize{X}{p}$, which denotes the multiset of connected components of FDDS~$X$ with cycles of length $p$, and the function~$\setDive{X}{p}$, denoting the multiset of connected components of $X$ with cycles of length \emph{dividing} $p$. In the rest of this section, we denote unroll trees with bold lowercase letters.

\begin{Lemma}\label{lemma:first_cond_tree}
	Let $P = \sum_{i=0}^{m} A_i X^i$ be a pseudo-injective polynomial and $B$ an FDDS.
	Let $c$ such that $\minLength{A_c} = \minLength{\sum_{i=1}^{m} A_i}$, let~$A = \sum_{i=1}^m A_i$ and let $\setDive{P}{\minLength{B - A_0}} = \sum_{i=1}^{m} \setDive{A_i}{\minLength{B - A_0}}$.
	If there exists $Y$ such that $P(Y) = B$ then $Y$ contains a connected component $C$ with minimal unroll tree equals to the minimal unroll tree give by the algorithm of \cite[Section 3]{poly_inj} with input $\setDive{P}{\minLength{B - A_0}}$ and $\setDive{B - A_0}{\minLength{B - A_0}}$ and cycle length $k \times \lcm(\aLcm{A_c}{B}, p ))$,
	where $p$ is the minimal period of $\tree{t}$ and $k>0$ is an integer in $\setDiv{\minLength{A_c}}$ such that $\gcd(k, \lcm(\aLcm{A}{B}, p ))= 1$.
\end{Lemma}

\begin{proof}
	We assume that there exists an FDDS $Y$ such that $P(Y) = B$.
	Then, there exists a connected component~$C$ of~$Y$ such that~$P(C)$ contains a connected component~$D$ of~$\setDive{B}{\minLength{B}}$ having~$\min(\unroll{D}) = \min(\unroll{\setDive{B}{\minLength{B}}})$.
	Hence~$P(Y) - A_0 = B'$ with $B' = B - A_0$.
	Then, since $\setDive{\cdot}{\minLength{B'}}$ is an endomorphism~\cite{poly_inj}, it follows that $\sum_{i=1}^{m} \setDive{A_i}{\minLength{B'}} (\setDive{Y}{\minLength{B'}})^i = \setDive{B'}{\minLength{B'}}$. 
	Thus, since unroll is a homomorphism $\mathbb{D} \to \mathbb{U}$ \cite{article_arbre}, we have
	\[
	\sum_{i=1}^{m} \unroll{\setDive{A_i}{\minLength{B'}}} \times (\unroll{\setDive{Y}{\minLength{B'}}})^i = \unroll{\setDive{B'}{\minLength{B'}}}.
	\]
	Consequently, there exists $i$ such that $\min(\unroll{\setDive{A_i}{\minLength{B'}}}) \min(\unroll{C})^i = \min(\setDive{B'}{\minLength{B'}})$.
	Therefore, the minimal unroll tree of $C$ is the minimal unroll tree give by the algorithm of \cite[Section 3]{poly_inj} with input $\setDive{P}{\minLength{B'}}$ and $\setDive{B'}{\minLength{B'}}$.
	
	Let $p$ be the period of $\min(\unroll{C})$.
	It follows that the cycle length~$c$ of~$C$ is a multiple of~$p$.
	By a reasoning similar to the proof of Lemma~\ref{lemma:first_cond_cycle_poly}, $c$ is a multiple of~$\aLcm{A}{B}$ and thus a multiple of~$\lcm(\aLcm{A}{B},p)$. Then, there exists a divisor~$k$ of~$\minLength{A}$ such that~$c = k \times \lcm(\aLcm{A}{B},p)$ with~$\gcd(k, \lcm(\aLcm{A}{B},p)=1$.
\end{proof}

Hence, we can easily generalize Lemma~\ref{lemma:second_cond_cycle_poly} like this :

\begin{Lemma}\label{lemma:second_cond_tree}
	Let $P = \sum_{i=0}^{m} A_i X^i$ be a pseudo-injective polynomial and $B$ an FDDS.
	Let $c$ such that $\minLength{A_c} = \minLength{\sum_{i=1}^{m} A_i}$ and let~$A = \sum_{i=1}^m A_i$.
	Then we have $P (Y + C ) = B$ if and only if $P (Y + d C') = B$ for all $d \in \setDiv{k}$, where $C$ and $C'$ are two connected components such that:
	\begin{itemize}
		\item $\min(\unroll{C}) = \min(\unroll{C'})$ and the period of this tree is $p$;
		\item $C$ has a cycle length of $k \times \lcm(\aLcm{A_c}{B-A_0}, p)$;
		\item $C'$ has a cycle length of $(k/d) \times \lcm(\aLcm{A_c}{B-A_0}, p)$.
	\end{itemize}
\end{Lemma} 

Thus, if we replace line~\ref{algo_poly_cycle:step2} of Algorithm~\ref{algo:poly_cycle} by:
\begin{enumerate}
	\item[3.1] Compute  $\setDive{P}{\minLength{B}}(X)~-~\setDive{P}{\minLength{B}}(Y) = \setDive{B}{\minLength{B}} + \setDive{P}{\minLength{B}}(Y)$ by the algorithm in \cite[Section 3]{poly_inj} and its minimal unroll tree in $\tree{x}$ if it exists; otherwise, abort the computation.
	\item[3.2] Let $p$ be the minimal period of $\tree{x}$.
	\item[3.3] Create $C$ the connected component whose cycle length is $\lcm(p, \aLcm{A_c}{B})$ such that $\tree{x} \in \unroll{C}$.
\end{enumerate}
we obtain a polynomial algorithm, since the new lines can also be executed efficiently. 
As before, the algorithm only returns correct solutions.
Proving that the algorithm does always find a solution when one exists requires a straightforward generalization of Lemma~\ref{lemma:correcteness_algo_div_cycle} (obtained by multiplying the value of~$\aLcm{A}{B}$ by~$p$ in the proof). This concludes the proof of Theorem~\ref{th:sol_poly_is_poly}.

Remark that a direct consequence of Lemmas~\ref{lemma:first_cond_tree} and~\ref{lemma:second_cond_tree} is that the equations considered in this section have a unique solution with the maximum number of connected components (which is returned by the three algorithms described above); this solution minimizes the lengths of the cycles. Symmetrically, there is a unique solution \emph{minimizing} the number of connected components.


	\section{Equations with cycles encoded compactly}

We will reexamine the cases of equations over sum of cycles, but with a different encoding. 
Indeed, although until now we have represented sums of cycles as explicit graphs (equivalent to a unary coding of the number of states in terms of size) in order to conserve the dynamical point of view, we can also represent them more compactly. 
Here we choose an encoding as arrays of pairs $(n,l)$ where~$l$ is the cycle length and $n$ is the number of copies of this cycle; if~$A$ is an FDDS encoded this way, we denote by~$A[i].l$ the~$i$-th cycle length, and by~$A[i].n$ its number of occurrences.
We will prove that equations over sums of cycles can be solved efficiently even under this coding, while the input size is sometimes reduced exponentially.

We introduce the following adaptation of Algorithm~\ref{algo:div_cycle}.

\begin{algo}\label{algo:div_cycle2}
	Given two sums of cycles $A$ and $B$ with $A$ pseudo-cancelable, we can compute $X$ such that $AX=B$, if any exists, in the following way:
	\begin{enumerate}
		\item Let $min = 0$.
		\item Sort $A$ and $B$ by length of the cycles.
		\item \label{algo_div_cycle2:step1} Check if the number of states of $A$ is less than that of $B$.
		\item\label{algo_div_cycle2:step2} Let $p = \aLcm{(B[min].l)}{A[0].l}$.
		\item \label{algo_div_cycle2:step3} Let $d$ be the number of cycles of length $B[min].l$ in the product $A C_p$.
		\item \label{algo_div_cycle2:step4} Let $S$ be sum of cycles such that $S.n = B[min].n / d$ and $S.l = p$. 
		\item\label{algo_div_cycle2:step5} Check if $AS$ is a subset of $B$.
		\item\label{algo_div_cycle2:step6} Add $S$ to $X$.
		\item\label{algo_div_cycle2:step7} Remove $AS$ from $B$ and set $min \gets \min \{ i \mid B[i].n > 0 \}$, if the set is nonempty, and go back to~\ref{algo_div_cycle2:step1}; otherwise, return $X$.   
	\end{enumerate}
\end{algo}

\begin{Lemma}
	Algorithm~\ref{algo:div_cycle2} is correct.
\end{Lemma}

\begin{proof}
	For reasons similar to Algorithm~\ref{algo:div_cycle}, if the algorithm returns a solution, then it is correct.
	For the other direction, if there exists a solution $X$ such that $AX = B$, then by Lemma~\ref{lemma:correcteness_algo_div_cycle} there exists a solution $Y = \sum_{i=1}^{m} k_i C_{\aLcm{\minLength{B_i}}{\minLength{A}}}$ where~$B_i$ is the value of~$B$ at the beginning of the $i$-th iteration.
	From Corollary~\ref{cor:fix_min}, it follows that, in $AY$, the term $A k_1 C_{\aLcm{\minLength{B_1}}{\minLength{A}}}$ contains all cycles of $B$ with length $\minLength{B}$. 
	Therefore $k_1$ is divisible by the number of cycles of $A C_{\aLcm{\minLength{B_1}}{\minLength{A}}}$ having length $\minLength{B}$. 
	By induction, the lemma follows.  
\end{proof}

Before the analysis of the complexity of Algorithm~\ref{algo:div_cycle2}, we define an efficient procedure for computing~$\aLcm{b}{a}$ for~$a,b>0$ with~$a$ dividing~$b$ (line~\ref{algo_div_cycle2:step2}), which does not require factoring either~$b$ or~$a$. This procedure is similar to Algorithm~8 in~\cite{riva2022thesis}.

\setcounter{algorithm}{\theTheorem}

\vspace{-1em}

\begin{algorithm}
	\caption{\texttt{computing} $\aLcm{b}{a}$}\label{algo:aLcm}
        \begin{multicols}{2}
	\begin{algorithmic}[1]
		\State $quot_1 \gets b / a$ \label{line:quot_1}
		\State $l_1 \gets \lcm(quot_1, a)$
		\State $quot_2 \gets b / l_1$
		\State $prod \gets a \times quot_2$
		\State $l_2 \gets \lcm(prod, b)$
		\State $res_1 \gets l_2 / b$
		\State $g \gets \gcd(b,prod)$
		\State $quot_3 \gets b / g$
		\State $quot_4 \gets quot_1 / quot_3$ \label{line:quot_4}
                \State $pow \gets quot_4^{\lceil \log_2 b \rceil}$ \label{line:pow}
                \State $res_2 \gets \gcd(b, pow)$ \label{line:res_2}
		\State \Return $res_1 \times res_2$
	\end{algorithmic}
        \end{multicols}
\end{algorithm}

\vspace{-1.5em}

\addtocounter{Theorem}{1}


\begin{figure}[t]
\centering
\includegraphics[page=1,width=\textwidth]{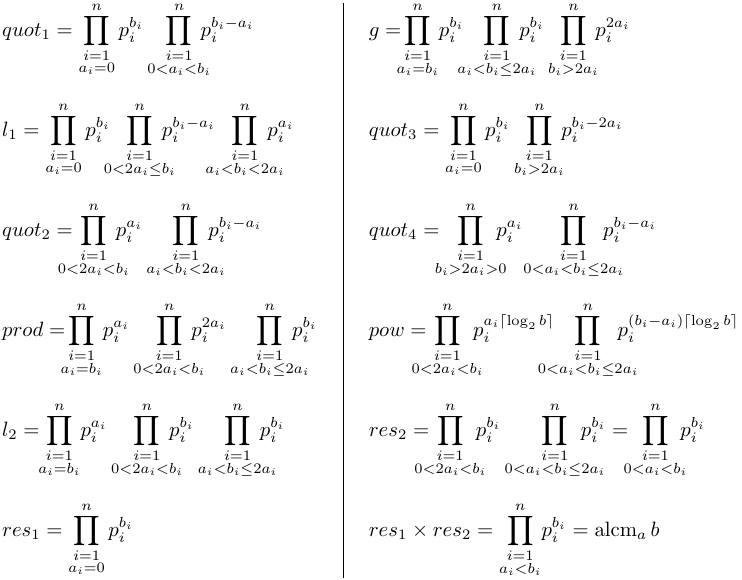}
\caption{The decomposition into primes of each variable occurring in Algorithm~\ref{algo:aLcm}, which proves the correctness of its output.}
\label{fig:alcm-calculations}
\end{figure}

 \begin{Lemma}\label{lemma:aLcm_poly}
 	Algorithm~\ref{algo:aLcm}~is correct and executes in $\bigo{(\log b)^2}$ time.
 \end{Lemma}
 
 \begin{proof}
 	Let $a =  \prod_{i=1}^{n} \qiai$ and $b = \prod_{i=1}^{n} \qibi$ with $p_i$ the $i$-th prime and large enough~$n$. 
 	Fig.~\ref{fig:alcm-calculations} shows the decomposition into primes of the variables of the algorithm; the terms of the factorization are grouped in order to make the calculations performed in each step more explicit.

        Line~\ref{line:res_2} of the algorithm is the only step that needs an argument more sophisticated than just an arithmetic computation in order to show that
        \[
        \gcd(b, pow) = \prodC{0 < 2a_i < b_i} \qibi \prodC{0 < a_i < b_i \le 2a_i} \qibi
        \]
        which implies that~$res_2$ gets the correct value for the computation of the result in the last line. Remark that~$\lceil \log_2 b \rceil \ge b_i$ for all~$i$. Indeed, $p_i^{b_i} \le b$ and since~$p_i \ge 2$, we have~$b_i \le b_i \log_2 p_i \le \log_2 b \le \lceil \log_2 b \rceil$. This implies~$a_i \lceil \log_2 b \rceil \ge b_i$ if~$a_i > 0$, which gives the first term of the product. For the second term, since~$b_i > a_i$, we have~$a_i \le b_i-1$ and thus~$(b_i - a_i) \lceil \log_2 b \rceil \ge \lceil \log_2 b \rceil \ge b_i$.

        

        In order to analyze the complexity of the algorithm, remark that the input integers are bounded by~$b$ and that each variable in lines~\ref{line:quot_1}--\ref{line:quot_4} is obtained by constant number of products, divisions, $\gcd$ and~$\lcm$; as a consequence, they are all bounded by a constant power~$c$ of~$b$. Assuming a constant-time model for basic arithmetic operations, the $\gcd$ and~$\lcm$ operations require time~$\bigo{\log b}$.
        In particular, this is the case for~$quot_4$: hence, $pow$ is computed in time~$\bigo{\log\log b}$ using fast exponentiation and is bounded by~$b^{c\lceil \log_2 \rceil}$, therefore its size is~$\bigo{(\log b)^2}$. Line~\ref{line:res_2} then requires~$\bigo{(\log b)^2}$ time, which gives the promised complexity bound.
\end{proof}

\begin{Lemma}\label{lemma:comp_div_cycle2}
	The complexity of Algorithm~\ref{algo:div_cycle2} is $\bigo{c^2 \times (\log c + (\log d)^2)}$ where $c$ is the maximum number of distinct cycle lengths of $A$ and $B$, and $d$ is the length of the longest cycle.
\end{Lemma}

\begin{proof}
	First of all, the number of iterations is bounded by the number of elements (distinct cycle lengths) of array $B$, hence by $c$. 
	Secondly, the product of two sums of cycles, where all cycle lengths of the second term are identical, can be performed in $\bigo{c^2 \log_2 d}$ time. 
	Indeed, for each pair of elements $C_1, C_2$, we just have to compute the $\gcd$ and the $\lcm$ of $C_1.l$ and $C_2.l$ and multiply the $\gcd$ by $C_1.n$ and $C_2.n$.
	This implies that line~\ref{algo_div_cycle2:step3} and the computation of $AS$ require $\bigo{c \log_2 d}$ time.
	Furthermore, since the number of elements of $AS$ is bounded by the number of elements of $A$ (since $S$ is only one element) and since $AS$ and~$B$ can be sorted, line~\ref{algo_div_cycle2:step7} and verifying if $AS$ is a subset of $B$ are accomplished in $\bigo{c}$ time. 
	Moreover, sorting $AS$ can be done in $\bigo{c \log_2 c}$ time, hence line~\ref{algo_div_cycle2:step5} is also $\bigo{c \log_2 c}$ time. 
	Additionally, line~\ref{algo_div_cycle2:step1} requires $\bigo{c}$ time, since it just an array traversal.
	Beside, by Lemma~\ref{lemma:aLcm_poly}, we execute line~\ref{algo_div_cycle2:step2} in $\bigo{(\log_2 d)^2}$ time. 
	Finally, all the other operations in the loop only require constant time, therefore the complexity of the loop is $\bigo{c^2 \log c + c^2 (\log_2 d)^2}$. 
	To conclude, since the operations  outside the loop require $\bigo{c \log c}$ time, the lemma follows.   
\end{proof}

As an aside, Lemma~\ref{lemma:comp_div_cycle2} has the consequence that, when the FDDS of the equation~$AX=B$ are sums of cycles represented explicitly as graphs, we can improve the runtime of Algorithm~\ref{algo:div_cycle} by first converting the input to the compact encoding described above.

\begin{Corollary}
	Let $A$ and~$B$ be sums of cycles with $A$ pseudo-cancelable.
	If $A$ and $B$ are given in input as graphs, then computing $X$ such that $AX = B$ can be performed in $\bigo{n (\log n)^2}$ time.
\end{Corollary}

\begin{proof}
	Translating a sum of cycles encoded as a graph into a sum of cycles encoded by a sorted array can be accomplished by a simple traversal, therefore in $\bigo{n \log n}$ in our context. 
	Remark that the number of elements $c$ of the array is $\bigo{\sqrt{n}}$. 
	Indeed, the smallest sum of cycles with $c$ different cycles has $\sum_{i=1}^{c} i = c(c+1)/2$ states. 
	At this point, by the Lemma~\ref{lemma:comp_div_cycle2}, we conclude that the complexity of finding the solution~$X$ is $\bigo{(\sqrt{n})^2 (\log \sqrt{n} + (\log n)^2)} = \bigo{n (\log n)^2}$ time.
\end{proof}

We can now generalize Lemma~\ref{lemma:comp_div_cycle2} to equations of the form~$P(X) = B$ with~$P$ a pseudo-injective polynomial and~$B$ a sum of cycles.
For this, we represent the polynomial $P = A_1 X^{p_1} + \cdots + A_m X^{p_m}$ (where the number of states of each $A_i$ is at least $1$) as an array of pairs $(p_i, A_i)$.

\begin{algo}\label{algo:poly_cycle2}
		Given a sum of cycles  $B$ and a pseudo-injective polynomial $P = \sum_{i=0}^{m} A_iX^i$ over sums of cycles encoded compactly, we can compute $Y$ such that $P(Y)=B$, if any exists, as follows:
		\begin{enumerate}
			\item Set $min \gets 0, B \gets B - A_0, P \gets P - A_0, A \gets \sum_{i=1}^{m} A_i$ and $X \gets \mathbf{0}$.
			\item Sort $A$ and $B$ by length of the cycles.
			\item \label{algo_poly_cycle2:step1} Check if $|A| \le |B|$. 
			\item\label{algo_poly_cycle2:step2} Set $l \gets \aLcm{(B[min].l)}{A[0].l}$, $d \gets 1$ and $d' \gets B[min].n$.
			\item \label{algo_poly_cycle2:step3} Set $a_{init} \gets P(Y)$ and $a_{end} \gets P(Y + d C_l)$ by computing progressively \linebreak $P(Y + dC_l)$ without exceeding the size of the original~$B$.
			\item Check if $d \neq d'$.
			\item \label{algo_poly_cycle2:step4} Sort the cycles of  $a_{init}$ and $a_{end}$ by length and set $rem\gets a_{init}-a_{end}$.   
			\item  \label{algo_poly_cycle2:step5}If $rem \nsubseteq B$ then set $d \gets \lfloor (d + d')/2 \rfloor$ and go back to~\ref{algo_poly_cycle2:step3}. Otherwise if $B[min].n - rem[0].n \neq 0$ then $d' \gets \lceil (d+ d')/2 \rceil$ and go back to~\ref{algo_poly_cycle2:step3}.
			\item\label{algo_poly_cycle2:step7} Otherwise add $(d, l)$ to $Y$.
			\item\label{algo_poly_cycle2:step8} Remove $rem$ from $B$ and set $min \gets \min \{ i \mid B[i].n > 0\}$, if the set is nonempty, and go back to~\ref{algo_div_cycle2:step1}; otherwise, return $Y$.   
		\end{enumerate}
\end{algo}

\begin{Theorem}
	Let $P$ be a pseudo-injective polynomial where each coefficient is a sum of cycles and $B$ a sum of cycles. 
	Then we can solve $P(X) = B$ in polynomial time.
\end{Theorem}

\begin{proof}
	Thanks to Proposition~\ref{prop:poly_cycle_polytime} and by applying the same reasoning as Lemma~\ref{lemma:aLcm_poly}, we can deduce that Algorithm~\ref{algo:poly_cycle2} only outputs valid solutions. 
	
	Assume~$B \ne \mathbf{0}$, $P= \sum_{i=0}^m A_iX^i$ and $P(Y) = B$ with $Y\neq \mathbf{0}$ and~$Y \ne \mathbf{1}$. Then, without loss of generality, we can choose $m \le \log_2 b$, where $b$ is the number of states in $B$. 
	Indeed, since $|Y^m| = |Y|^m$, we have~$2^m \le |Y^m| \le b$ and thus~$m \le \log_2 b$.
	Observe that we can enumerate all natural numbers between $0$ and $m$ in polynomial time; indeed, the size of the representation of $B$ is $\sum_{i=0}^{n-1} (\log_2 B[i].n + \log_2 B[i].l) = \log_2 \prod_{i=0}^{n -1} (B[i].n \times B[i].l)$ with $n=|\setLength{B}|$, and $ \log_2 b = \log_2 \sum_{i=0}^{n-1} (B[i].n \times B[i].l)$. 
	
	Let $X = \sum_{i=1}^k C_{x_i}$ be a sum of cycles and $j\ge 1$ be an integer. 
	Then $C_x^j$ is an element of $X^j$ for each $x \in \setLength{X}$. 
	Since $C_p^q = p^{q-1} C_p$ for all positive integers $p,q$, we deduce that 
        $X^i \subseteq X^{i+1}$ for all integer $i \ge 0$ and thus, for all $A$, we have $AX^i \subseteq AX^{i+1}$. 
	Thanks to the two previous observations, line~\ref{algo_poly_cycle2:step3} can be executed in polynomial time.
	Since the number of iterations of Lines~\ref{algo_poly_cycle2:step3}--\ref{algo_poly_cycle2:step5} perform a binary search ranging from $1$ to $B[min].n$ and thus require at most $\log_2 (B[min].n)$ iterations, this loop can be carried out in polynomial time.
	
	Finally, since the number of iterations of lines~\ref{algo_poly_cycle2:step1}--\ref{algo_poly_cycle2:step8} is bounded by $|\setLength{B}|$, the statement of the theorem follows.
\end{proof}


	\section{Conclusions}

In this paper we enlarged the class of efficiently solvable polynomial equations over FDDS by generalizing the notion of injectivity of polynomials. Furthermore, we developed a procedure that allows us to carry over these results to a more compact representation when the FDDS involved are sums of cycles, not only without incurring in an exponential slowdown, but even improving the efficiency of one of the aforementioned algorithms.

This work also highlighted the structure of solutions to equations over pseudo-injective polynomials (Lemmas~\ref{lemma:first_cond_tree} and~\ref{lemma:second_cond_tree}), which we expect will be useful for efficiently counting and enumerating the set of solutions. We also believe that a similar structure might be found among the solutions of other equations of the form~$P(X)=B$, if the conditions on~$P$ are relaxed but other conditions on the divisibility of the lengths of the cycles of~$B$ are introduced.

A more in-depth investigation of compact representations of FDDS in the context of polynomial equations is also interesting, since in most applications the dynamics of the systems are not described explicitly, and the detection of simple dynamical properties usually becomes intractable, notably with a succinct encoding by Boolean circuits.

Another open problem is finding complexity lower bounds for different classes of equations. The exact complexity of solving equations of the form~$AX=B$ or, more generally, $P(X)=B$ of a single variable, but without other restrictions, is still open. Is it an~$\NP$-hard problem or does it admit an efficient algorithm? Does this generalize to any constant number of variables? Furthermore, how much can we relax the constraints on the number of variables and the fact that the right-hand side of equations must be constant without them becoming intractable or even undecidable~\cite{article_fondateur}?


        \begin{credits}
        \subsubsection{\ackname}
        This work has been partially supported by the HORIZON-MSCA-2022-SE-01 project 101131549 ``Application-driven Challenges for Automata Networks and Complex Systems (ACANCOS)'' and by the project ANR-24-CE48-7504 ``ALARICE''.
	\end{credits}
	
	\bibliography{biblio}
	\bibliographystyle{unsrt}
	
\end{document}